\documentclass[twocolumn,10pt]{article}

\usepackage[margin=0.75in]{geometry}
\usepackage{graphicx,amsmath,bm}
\usepackage{booktabs}
\usepackage{balance}
\usepackage{authblk}              

\usepackage{xcolor}
\definecolor{mygray}{gray}{0.45}  

\usepackage[superscript]{cite}
\usepackage[
  colorlinks   = true,
  citecolor    = mygray,
  linkcolor    = mygray,
  urlcolor     = mygray]{hyperref}

\makeatletter
\renewcommand{\tagform@}[1]{%
  \textcolor{mygray}{(\ignorespaces#1\unskip)}}
\makeatother

\title{\textbf{From Density Functional Theory to Spin Hamiltonians:\\
Magnetism in $d^5$ Honeycomb Compound OsCl$_3$}}

\author{Ritwik Das}
\author{Indra Dasgupta%
  \thanks{E-mail: %
    \href{mailto:intrd@iacs.res.in}{intrd@iacs.res.in}, %
    \href{mailto:sspid@iacs.res.in}{sspid@iacs.res.in}}}

\affil{School of Physical Sciences, Indian Association for the Cultivation
of Science\protect\\
2A and 2B Raja S.\ C.\ Mullick Road, Jadavpur, Kolkata 700 032, India}

\date{}  
\begin{document}

\twocolumn[ 
\maketitle

\begin{abstract}
Magnetism in strongly correlated honeycomb systems with $d^5$ electronic configuration has garnered significant attention due to its potential to realize the Kitaev spin liquid state, characterized by exotic properties. However, real materials exhibit not only Kitaev exchange interactions but also other magnetic exchanges, which may drive the transition from a spin liquid phase to a long-range ordered ground state. This work focuses on modelling the effective spin Hamiltonian for two-dimensional (2D) honeycomb magnetic systems with $d^5$ electronic configurations. The Hubbard-Kanamori (HK) Hamiltonian equipped with spin-orbit coupling and electron correlations is considered where onsite energies and hopping parameters, preserving the crystal symmetry, are extracted from the first principles Density functional theory (DFT) calculations. Exact diagonalization (ED) calculations for the HK Hamiltonian on a two-site cluster are performed to construct the effective magnetic Hamiltonian. The ground-state magnetic properties are explored using the semi-classical Luttinger-Tisza approach. As a representative case, the magnetic ground state of the $d^5$ honeycomb system OsCl$_3$ is investigated, and the variation of magnetic exchange parameters with respect to the correlation strength \( U \) and Hund's coupling \( J_H \) is analyzed. The magnetic ground state exhibits zigzag antiferromagnetic ordering for a chosen value of $U$ and $J_H$, consistent with DFT results. This study provides insight into the magnetism of OsCl$_3$ and offers a computationally efficient alternative to traditional energy-based methods for calculating exchange interactions for strongly correlated systems.
\end{abstract}

\vspace{1em}

\vspace{1.5em}
] 

\section{Introduction}

The magnetism in strongly correlated honeycomb systems with $d^5$ electronic configurations has attracted significant interest due to its potential to realize the novel Kitaev spin liquid state, known for its exotic properties \cite{Kitaev_1, Kitaev_2}. In practice, the magnetism in $d^5$ systems is not only dictated by Kitaev exchange interactions but also other magnetic exchanges, which may drive the transition from the spin liquid phase to a long-range ordered ground state in real materials \cite{Kitaev_3, Kitaev_4, Kitaev_5}.

This work focuses on modeling the effective spin Hamiltonian for two-dimensional (2D) magnetic systems, with particular emphasis on $d^5$ compounds in a honeycomb lattice. To study these systems, density functional theory (DFT) calculations, combined with wannierization, are employed to derive various hopping parameters between maximally localized Wannier functions (MLWFs), preserving the underlying structural symmetry of the material. To account for electron-electron interactions and spin-orbit coupling, the Hubbard-Kanamori Hamiltonian retaining only the t$_{2g}$ states is employed, which captures the complex interplay between these effects in the low-energy regime. Exact diagonalization (ED) calculations on a two-site cluster are then performed to extract the effective magnetic Hamiltonian, providing insight into the nature of magnetic exchanges within the system. The ground-state magnetic properties are explored using the semiclassical Luttinger-Tisza approach, which enables the determination of the magnetic ground state.

\begin{figure}[!t]
\centering{
\includegraphics[width=1\columnwidth]{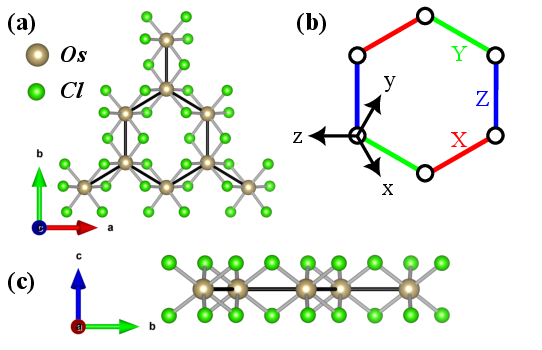}
}
\caption{Crystal structure of OsCl$_3$: \textbf{(a)} Os atoms form a honeycomb network, with Cl ligands creating an octahedral environment. \textbf{(b)} Nearest-neighbour bonds X (red), Y (green), Z (blue), and the local coordinate system (x, y, z) defined by Cl ligands, forming an octahedral cage around Os. \textbf{(c)} Alternative viewpoint of the OsCl$_3$ structure.}\label{fig1}
\end{figure}

In particular, in this work, the magnetic ground state of the honeycomb compound OsCl$_3$ \cite{OsCl3_2} is investigated, focusing on the variation of magnetic exchange parameters with respect to the correlation strength \( U \) and Hund's coupling \( J_H \). The magnetic ground state for a chosen value of $U$ and $J_H$ relevant for OsCl$_3$ is found to exhibit zigzag antiferromagnetic ordering, which is in agreement with the results obtained from ab-initio DFT calculations. This study provides valuable insights into the magnetism of OsCl$_3$ and offers a computationally efficient alternative to traditional energy-based methods for calculating exchange interactions in strongly correlated systems.

The paper is organized as follows: In Section 2, the crystal structure and non-spin polarized DFT calculations are presented, along with a detailed symmetry analysis of the structure and its impact on crystal field and nearest-neighbor hopping parameters. These parameters are extracted using the wannierization technique. Section 3 outlines the method for extracting the low-energy effective spin Hamiltonian using ED method for a two site cluster for the description of magnetism. In Section 4, the semi-classical Luttinger-Tisza approach is applied to predict the magnetic ground state of the OsCl$_3$ system. Finally, section 5 presents the conclusion.

\section{Crystal Structure, Computational Details and Non-spinpolarized Electronic Structure}

The crystal structure of OsCl$_3$ (C2/m space group, No. 12) is shown in Fig.\ref{fig1}\cite{OsCl3_1, OsCl3_2, OsCl3_3}. As seen in Fig.\ref{fig1}(a) and (c), Os atoms form a 2D honeycomb network, while Cl ligands create a distorted octahedral environment. Fig.\ref{fig1}(b) illustrates the local coordinate system $(x,y,z)$, used for calculating the electronic band structure and partial density of states (DOS) using DFT. The nearest-neighbour bonds X, Y, and Z in the honeycomb network are also highlighted (see Fig.\ref{fig1}(b)).

\begin{figure}[!t]
\centering{
\includegraphics[width=1\columnwidth]{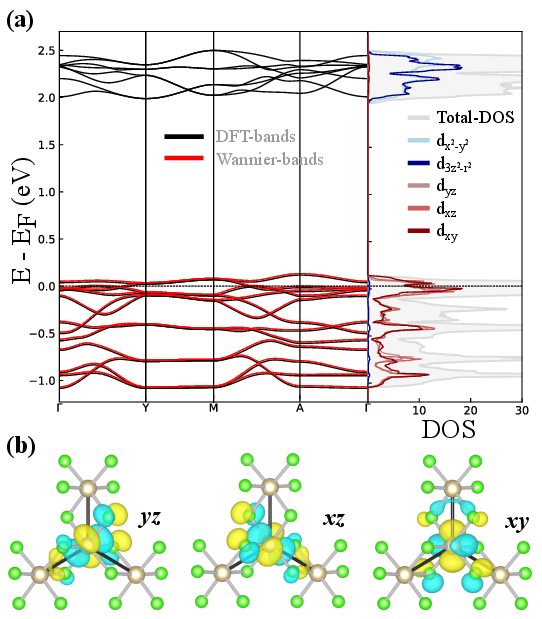}
}
\caption{Electronic properties: \textbf{(a)} Non-spinpolarized electronic band structure and DOS, showing the dominance of the $d$-$t_{2g}$ bands of the Os atom near $E_F$. \textbf{(b)} $t_{2g}$ MLWFs obtained from Wannierization, oriented according to the local coordinate system.}\label{fig2}
\end{figure}

\begin{table*}[h!]
\centering
\setlength{\tabcolsep}{6pt} 
\renewcommand{\arraystretch}{1.5} 
\begin{tabular}{|c|c|c|c|c|c|c|c|c|c|c|c|c|}
    \hline
    $\Delta_1$ & $\Delta_2$ & $\Delta_3$ & $t_1$ & $t_{1a}'$ & $t_{1b}'$ & $t_2$ & $t_2'$ & $t_3$ & $t_3'$ & $t_4$ & $t_{4a}'$ & $t_{4b}'$ \\
    \hline
    -31.2 & -22.7 & -18.1 & 66.1 & 51.9 & 57.1 & 186.1 & 189.0 & -201.9 & -141.9 & -18.9 & -20.5 & -18.7 \\
    \hline
\end{tabular}
\caption{Nonmagnetic hopping parameters for $OsCl_3$ in meV.}
\label{nonmag data}
\end{table*}

DFT calculations were performed using the plane-wave-based projector augmented wave (PAW) method as implemented in the Vienna ab initio simulation package (VASP) with the generalized gradient approximation for exchange-correlation \cite{DFT_1, DFT_2, DFT_3}. The plane-wave cutoff was set to 500 eV, and a $\Gamma$-centered $9 \times 9 \times 3$ k-point mesh was used for Brillouin zone (B.Z.) integration.

The non spin polarized electronic band structure and density of states (DOS) obtained from DFT calculations are shown in Fig.\ref{fig2}(a). The calculations, using the local coordinate system defined in Fig.\ref{fig1}(b), reveal significant crystal field splitting, with the Os $d$-$t_{2g}$ bands dominating near the Fermi level ($E_F$). 

The Os $d$-electronic levels are split into a three-fold degenerate $t_{2g}$ and a two-fold degenerate $e_g$ configuration due to the octahedral crystal field symmetry ($O_h$) provided by the Cl-ligands (Fig.\ref{fig1}(a)). For the $d^5$ configuration, only the $t_{2g}$ levels are relevant. However, as the Os atoms have the site symmetry $C_2$ in the C2/m space group, the $t_{2g}$ levels undergo further splitting. For OsCl$_3$, the C$_2$ axis is along the Z-bond.

To extract the hopping parameters for the low energy tight binding model, we applied the wannierization technique using the code Wannier90 \cite{wan90_1, wan90_2}. Maximally localized wannier functions (MLWFs) were derived from the d-$t_{2g}$ projections of Os atoms, and the resulting MLWFs are displayed in Fig.~\ref{fig2}(b).

In the tight-binding model for the non-interacting system, hopping terms between Os sites derived from DFT calculations are included. Due to crystal symmetry, the elements of these hopping matrices are constrained. In the honeycomb structure, the three nearest-neighbor bonds are denoted as X, Y, and Z, aligned with the local $x$, $y$, and $z$ axes, respectively (Fig.\ref{fig1}(b)). Based on these symmetry considerations, the crystal field matrix for the $t_{2g}$ levels and the nearest-neighbor hopping matrices are given by \cite{projED_1}:

\[
H_{CF} = 
    \left( \begin{array}{ccc}
    0 & \Delta_1 & \Delta_2 \\
    \Delta_1 & 0 & \Delta_2 \\
    \Delta_2 & \Delta_2 & \Delta_3
    \end{array} \right)
	\quad
	T_1^{Z} = 
	\left( \begin{array}{ccc}
    t_1 & t_2 & t_4 \\
    t_2 & t_1 & t_4 \\
    t_4 & t_4 & t_3
    \end{array} \right)
\]

\[
\begin{array}{cc}
T_1^{X} = 
\left( \begin{array}{ccc}
    t_3' & t_{4a}' & t_{4b}' \\
    t_{4a}' & t_{1a}' & t_2' \\
    t_{4b}' & t_2' & t_{1b}'
    \end{array} \right)
    & 
    T_1^{Y} = 
    \left( \begin{array}{ccc}
    t_{1a}' & t_{4a}' & t_2'  \\
    t_{4a}' & t_3' & t_{4b}' \\
    t_2' & t_{4b}' & t_{1b}'
    \end{array} \right)
\end{array}
\]

The above matrices are written with respect to the t$_{2g}$ basis maintaining the order ${yz,xz,xy}$. The corresponding values of these parameters are listed in Table \ref{nonmag data}.

\section{Determination of Magnetic Exchange Parameters}

In this section, we shall extract the magnetic exchange parameters using the multi-orbital Hubbard-Kanamori Hamiltonian with spin-orbit coupling, where the crystal field and hopping parameters are derived from DFT calculations for the t$_{2g}$ states:
\begin{equation}
H_{tot} = H_{CF} + H_{hop} + H_{SOC} + H_{U}
\end{equation}

The spin-orbit coupling (SOC) term is:

\begin{equation}
H_{SOC} = \lambda \vec{L}\cdot\vec{S} = \sum_i \lambda_i \vec{l}_i \cdot \vec{s}_i
\end{equation}

$\lambda$ is the SOC strength. The interaction term is given by:

\begin{equation}
\begin{split}
H_U = U \sum_{i, a} n_{i,a,\uparrow} n_{i,a,\downarrow} + (U' - J_H)\sum_{i, a<b,\sigma} n_{i,a,\sigma} n_{i,b,\sigma} \\
+ U' \sum_{i, a \neq b} n_{i,a,\uparrow} n_{i,b,\downarrow} - J_H \sum_{i, a \neq b} c_{i,a,\uparrow}^{\dagger} c_{i,a,\downarrow} c_{i,b,\downarrow}^{\dagger} c_{i,b,\uparrow} \\
+ J_H \sum_{i, a \neq b} c_{i,a,\uparrow}^{\dagger}c_{i,a,\downarrow}^{\dagger} c_{i,b,\downarrow} c_{i,b,\uparrow}
\end{split}
\end{equation}

Here $U$ is the intra-orbital Hubbard parameter and $J_H$ is the Hund's coupling. $U' = U - 2J_H$ is the inter-orbital Hubbard interaction strength.

To extract the low-energy effective spin Hamiltonian, we first consider the single-site case. Inclusion of the interaction term \( H_U \), the system, with five electrons (or one hole) in the \( t_{2g} \) levels, has six possible configurations (\( ^6C_5 = 6 \)), resulting in a six-fold degenerate ground-state manifold.

Incorporating \( H_{SOC} \), this degeneracy splits into a two-fold degenerate \( j = 1/2 \) ground state and a four-fold \( j = 3/2 \) excited state. The two-fold degenerate \( j=1/2 \) ground states form Kramer's doublets in the absence of crystal field distortions and behave as pseudo-spin-1/2 states. These pseudo-spins are the low-energy degrees of freedom for $d^5$ strongly correlated system that arise from the multiplet degeneracy rather than real electron spin.

\begin{figure}[!t]
\centering{
\includegraphics[width=1\columnwidth]{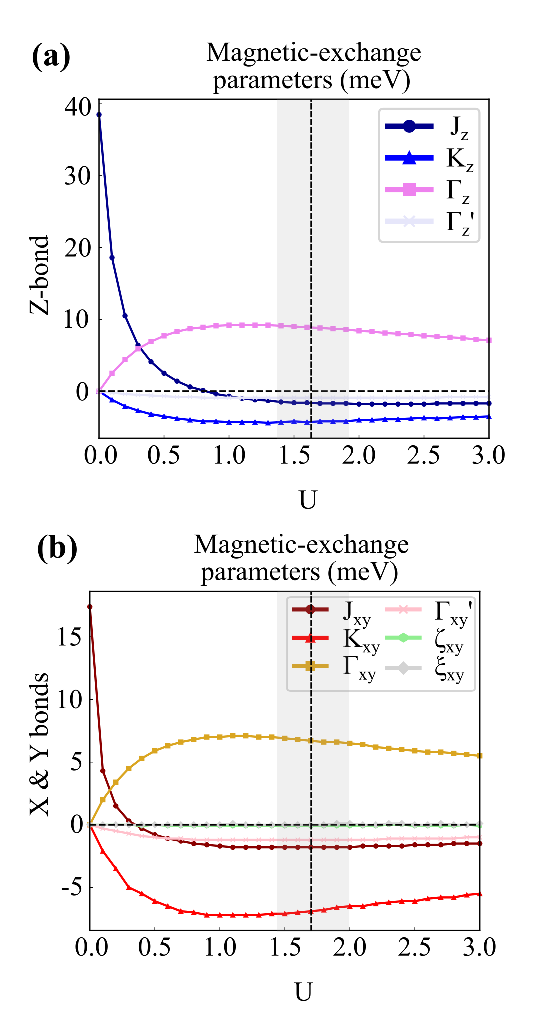}
}
\caption{Variation of magnetic-exchange parameters with correlation parameter $U$ for first nearest-neighbour bonds: \textbf{(a)} Z-bond and \textbf{(b)} X and Y bonds, for $J_H/U = 0.18$.}\label{fig3}
\end{figure}

Using the eigenstates of \( H_U + H_{SOC} \) from the single-site case, we now extend the system to two sites. In the absence of crystal-field and hopping terms, the two-site ground state is of the form \( |j_1=1/2, j_{1z} \rangle \otimes |j_2=1/2, j_{2z} \rangle \), resulting in a four-fold degenerate ground-state manifold.

When \( H_{CF} \) is included for the single site and \( H_{CF} + H_{hop} \) for the two-site system, the Kramer's degeneracy is lifted, breaking the four-fold degeneracy. However, this ground-state manifold remains well-separated from the excited states, and the exact eigenspectrum of the ground state projects onto the \( |j_1=1/2, j_{1z} \rangle \otimes |j_2=1/2, j_{2z} \rangle \) manifold. Given the large gap between the ground state and the lowest excited state, we focus on the many-body ground-state manifold.

To construct the low-energy Hamiltonian\cite{projED_1, projED_2, projED_3}, we numerically project the exact ground-state manifold onto the \( |j_1=1/2, j_{1z} \rangle \otimes |j_2=1/2, j_{2z} \rangle \) basis, ignoring the \( H_{CF} \) and \( H_{hop} \) terms. This yields a \( 4 \times 4 \) matrix in the pseudo-spin basis for the description of magnetism.

In this approach, ED results for the two-site system are used to construct the low energy effective spin Hamiltonian. The only approximation is the projection onto the ground-state manifold, which is numerically justified due to the large energy gap. While perturbation theory would similarly involve projection, it requires an expansion depending on U, J$_H$ and hopping parameters, the proposed method however is more accurate, as it uses the exact spectrum obtained from two-site ED.

The magnetic exchange parameters for the low-energy effective Hamiltonian follow the same symmetry considerations as the crystal field and hopping parameters. Therefore, the magnetic exchange Hamiltonian, expressed in terms of spin components \( S_i^x, S_i^y, S_i^z \) at site \( i \) and \( S_j^x, S_j^y, S_j^z \) at its nearest neighbor \( j \), depends on the specific nearest-neighbor bond (X, Y, or Z) and takes the form:

\[
J_1^{Z} = 
\renewcommand{\arraystretch}{1.5}
\begin{array}{c@{\hspace{0.3cm}}c@{\hspace{0.8cm}}c}
    &
    \left( \begin{array}{ccc}
    J_1^z & \Gamma_1^z & \Gamma_1^{'z} \\
    \Gamma_1^z & J_1^z & \Gamma_1^{'z} \\
    \Gamma_1^{'z} & \Gamma_1^{'z} & J_1^z + K_1^z
    \end{array} \right)
\end{array}
\]
\[
J_1^{X} = 
\renewcommand{\arraystretch}{1.5}
\begin{array}{c@{\hspace{0.3cm}}c@{\hspace{0.8cm}}c}
    &
    \left( \begin{array}{ccc}
    J_1^{xy} + K_1^{xy} & \Gamma_1^{'xy} + \zeta_1 & \Gamma_1^{'xy} - \zeta_1 \\
    \Gamma_1^{'xy} + \zeta_1 & J_1^{xy} +\xi_1 & \Gamma_1^{xy} \\
    \Gamma_1^{'xy} - \zeta_1 & \Gamma_1^{xy} & J_1^{xy} -\xi_1
    \end{array} \right)
\end{array}
\]
\[
J_1^{Y} = 
\renewcommand{\arraystretch}{1.5}
\begin{array}{c@{\hspace{0.3cm}}c@{\hspace{0.8cm}}c}
    &
    \left( \begin{array}{ccc}
    J_1^{xy} +\xi_1 & \Gamma_1^{'xy} + \zeta_1 & \Gamma_1^{xy}  \\
    \Gamma_1^{'xy} + \zeta_1 & J_1^{xy} + K_1^{xy} & \Gamma_1^{'xy} - \zeta_1 \\
    \Gamma_1^{xy} & \Gamma_1^{'xy} - \zeta_1 & J_1^{xy} -\xi_1
    \end{array} \right)
\end{array}
\]

Here, $J_1^z$ and $J_1^{xy}$ are the Heisenberg exchange parameters along the Z and X/Y nearest-neighbor bonds, respectively. The Heisenberg exchange parameters remain the same along the X and Y bonds, following the symmetry argument mentioned above. Similarly, the Kitaev exchange parameter is denoted by $K_1^z$ along the Z bond and $K_1^{xy}$ along the X and Y bonds. Other anisotropic terms such as $\Gamma$, $\Gamma'$, $\zeta$, and $\xi$ arise from the interplay between spin-orbit coupling (SOC) and electronic correlation effects. Again, due to symmetry, these terms take equal values along the X and Y bonds but differ along the Z bond.

The variations of the magnetic exchange parameters with respect to \( U \) are shown in Fig.\ref{fig3}(a) and (b) for the Z and X,Y bonds, respectively. For these variations, the ratio \( J_H/U = 0.18 \) is kept fixed, and the possible physical range for OsCl$_3$ is indicated by the shaded region in Fig.\ref{fig3}(a) and (b), where the magnetic exchange parameters remain nearly constant. Our calculations reveal that nearest neighbour Heisenberg and Kitaev exchanges are ferromagnetic throughout the possible physical range of U and J$_H$, however the antiferromagnetic pseudodipolar interaction parameters $\Gamma$ dominates so the system is likely to display long range magnetic order.

\section{Ground State Magnetic Order: The Luttinger-Tisza Approach}

We now obtain the ground-state magnetism using the semi-classical Luttinger-Tisza approach\cite{LT_0,LT_1}. Considering the spin Hamiltonian:

\begin{equation}
\label{eq: Hamiltonian}
E = \sum_{\substack{i,m,\alpha \\ j,n,\beta}} J_{i,m;j,n}^{\alpha,\beta} S_{i,m}^{\alpha} S_{j,n}^{\beta}
\end{equation}

where \( i, j \) are the unit cell indices (abbreviations for \( \vec{r}_i \) and \( \vec{r}_j \)), and \( m, n \) are the basis indices (there are two basis attached per unit cell for honeycomb structure). The exchange parameters \( J_{i,m;j,n}^{\alpha,\beta} \) have been calculated for the OsCl$_3$ system. Instead of treating the spin components as quantum mechanical operators (following the angular momentum algebra for spin-\( 1/2 \)), we treat the spin components classically in this approach. Due to translational symmetry, we can apply a Fourier transformation with respect to the unit cell indices to move to \( k \)-space:

\begin{equation}
E = \sum_{\vec{k} \in \text{BZ}} S^{\dagger}(\vec{k}) \cdot J(\vec{k}) \cdot S(\vec{k}) = \sum_{\vec{k} \in \text{BZ}} H(\vec{k})
\end{equation}

After diagonalization, the Hamiltonian in k-space, it can be expressed as:
\begin{equation}
E = \sum_{\vec{k} \in \text{BZ}} \sum_{\mu} \lambda^{\mu}(\vec{k}) |c^{\mu}(\vec{k})|^2
\end{equation}

where \( \lambda^{\mu}(\vec{k}) \) are the eigenvalues of \( J(\vec{k}) \), representing the energies associated with different spin modes at each \( \vec{k} \).

\begin{figure}[!t]
\centering{
\includegraphics[width=1\columnwidth]{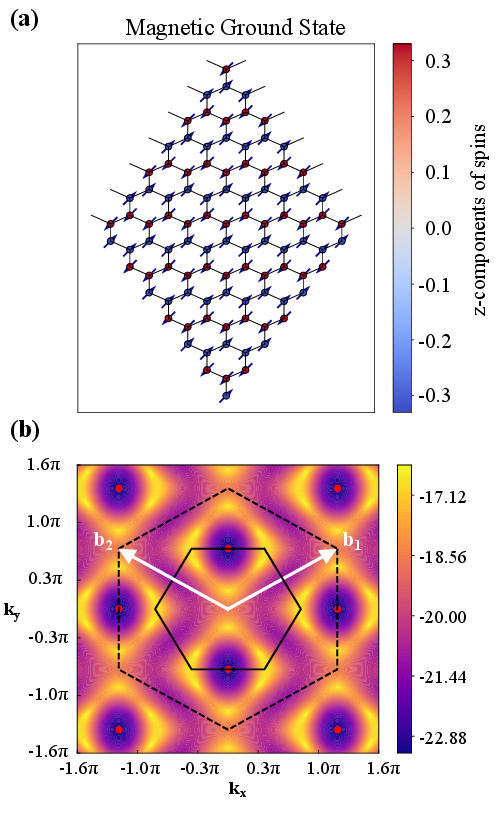}
}
\caption{Semi-classical ground state obtained from Luttinger-Tisza method:\textbf{(a)} The magnetic ground state of OsCl$_3$. \textbf{(b)} The minimum energy distribution obtained from Luttinger-Tisza approach.}\label{fig4}
\end{figure}

To determine the classical ground state configuration, we select the minimum value of \( \lambda^{\mu}(\vec{k}) \) with respect to \( \mu \) and \( \vec{k} \). If the minimum occurs at a particular wave-vector \( \vec{Q} \) within the Brillouin zone, then due to the symmetry of the B.Z., we must consider both \( \vec{Q} \) and \( -\vec{Q} \), ensuring \( \tilde{S}(\vec{Q}) = \tilde{S}(-\vec{Q}) \). For \( \vec{k} \neq \pm\vec{Q} \), we set \( \tilde{S}(\vec{k}) = 0 \).

Since \( |S(\vec{Q})|^2 = |S(-\vec{Q})|^2 \), the total energy can be expressed as:

\begin{equation}
E = 2 \lambda_{\text{min}}(\vec{Q}) |S(\vec{Q})|^2
\end{equation}

This choice must satisfy the weak constraint condition, ensuring that the total spin magnitude across all sites remains consistent with the required value. According to the weak condition, the magnitude of the total spin is given by:

\begin{equation}
\sum_{i,m,\alpha}\left[S_{i,m}^{\alpha}\right]^{*} S_{i,m}^{\alpha} = NZS_0^2
\end{equation}

where \( Z \) is the number of basis vectors per unit cell, and \( N \) is the number of unit cells. This condition does not ensure that the spin magnitude at each site is exactly \( S_0^2 \) (the strong condition), but as we only constrain the weak condition, this method is semi-classical, allowing fluctuations at different sites in reciprocal space while seeking classical ground state configurations.

Using the Luttinger-Tisza method \cite{LT_2, LT_3}, we obtain the magnetic ground state as a zigzag antiferromagnetic phase for $U=1.7$ eV and $J_H=0.3$ eV, as shown in Fig.\ref{fig4}(a). The corresponding \( \vec{Q} \), for which the energy of the spin system is minimized, is shown in Fig.\ref{fig4}(b). This result is consistent with the ab-initio DFT+U calculations\cite{OsCl3_2}.

\section{Conclusion}

In this work, we have studied the ground-state magnetism of the honeycomb system OsCl$_3$. We first extracted the crystal field splitting and hopping parameters for OsCl$_3$ using DFT and wannierization calculations. These parameters were employed to construct multi-orbital Hubbard-Kanamori model that incorporates correlation effects along with spin-orbit coupling. Further ED calculations of the Hubbard-Kanamori model defined on a two-site dimer for each bond was employed to calculate the exchange interactions and construct the low energy effective spin Hamiltonian.

We have analyzed the variation of the magnetic exchange parameters with respect to \( U \) and \( J_H \). In the parameter range physically relevant for OsCl$_3$, the Heisenberg and Kitaev exchanges are found to be ferromagnetic in nature, but the appreciable antiferromagnetic pseudodipolar interaction parameter $\Gamma$ favor long range magnetic order. The semi-classical Luttinger-Tisza approach was employed to obtain the ground state of the resulting magnetic Hamiltonian. The ground state is zigzag-antiferromagnetic in nature in agreement with DFT calculations available in the literature\cite{OsCl3_2}. Our analysis provides insight into the origin of magnetism in OsCl$_3$ and the magnetic ground state. Additionally, our method offers an alternative to energy-based methods for calculating magnetic exchange parameters for such systems using DFT.

\section*{Acknowledgement}
I.D thanks Prof. Swapan K Ghosh for introducing him to the beautiful and elegant subject of density functional theory. R.D thanks the Council of Scientific and Industrial Research (CSIR), India for research fellowship (File No. 09/080(1171)/2020-EMR-I). I.D would like to thank the Science and Engineering Research Board (SERB) India (Project No. CRG/2021/003024) and Technical Research Center, Department of Science and Technology Government of India for support.

\bibliographystyle{unsrt}  
\bibliography{references}   

\end{document}